\documentclass[12pt]{article}

\usepackage{graphicx}
\usepackage{multirow}
\usepackage{amsmath}
\usepackage{natbib}
\usepackage{xcolor}
\usepackage{enumitem}

\usepackage[title]{appendix}
\usepackage[affil-sl]{authblk}

\usepackage[colorlinks=true,allcolors=blue]{hyperref}

\begin{document}

\title{Impact of an Employment Policy on Companies' Expectations Fulfilment}

\author[1]{Javier Espinosa Brito}
\author[2]{Carlos Yévenes Ortega}
\author[3]{Gonzalo Franetovic Guzmán}
\author[4]{Diana Ochoa Diaz}
\affil[1,2]{Department of Economics, Universidad de Santiago de Chile\\ }
\affil[3,4]{Insitituto Nacional de Estadísticas de Chile 
}

\date{}

\maketitle

\begin{abstract}

We study the effect of Chile’s Employment Protection Law (Ley de Protección del Empleo, EPL), a law which allowed temporal suspensions of job contracts in exceptional circumstances during the COVID-19 pandemic, on the fulfillment of firms’ expectations regarding layoffs.  We use monthly surveys directed at a representative group of firms in the national territory. This panel data allows to follow firms through time and analyze the match between their expectations and the actual realization to model  their expectation fulfilment. We model the probability of expectation fulfilment through a logit model that allows for moderation effects. Results suggest that for those firms that expected to fire workers, for the firms that used the EPL, the odds they finally ended up with a job separation are 50\% of the odds for those that did not used the EPL. Small firms increase their probability of expectation fulfilment in 11.9\% when using the EPL compared to large firms if they declared they were expecting to fire workers.

\bigskip
{\bf Keywords:} Expectations fulfillment, Employment, Economic Policy.

%{\bf MSC2010:} xxx, yyy, zzz.

\end{abstract}

\section{Introduction} \label{intro}

Many policies arose during the onset of the COVID-19 pandemic in order to protect economic activities and employment. Long quarantines, curfews, capacity restrictions threatened with producing one of the deepest economic crises the world had seen. 

 The ILO Monitor \cite{ilo_ilo_2021} reports that in 2020 the world lost 114 million jobs relative to 2019 of which 33 million shifted to unemployment and 81 million shifted to inactivity. Furthermore, employment losses were highest in the Americas. 
 Chile was not the exception. The drop in employment during the June-August 2020 quarter was around 20\% which brought about an increase in the unemployment rate and a strong leaving of workers from the labor force resulting in around one out of three people of working age got unemployed or part of the potential labor force \cite{montt_chile_nodate}.
 
 Conventional economic policies such as stimulating aggregate demand or providing loans to companies showed a low impact in employment during the COVID-19 crisis \citep{chetty_economic_2020}. \cite{chetty_economic_2020} point out that transfers to low-income households evidenced a low impact on employment in the businesses most affected during 2020. Besides, they showed evidence that Paycheck Protection Program loans increased employment by about 2\% in small businesses. \cite{guerrieri_macroeconomic_2021} set up a model with multiple sectors where supply shocks could generate changes in aggregate demand even larger than the shocks themselves. In their model, as some sectors were shut down, conventional fiscal stimulus was not so effective but social insurance policy made it easier to achieve the demand stabilization objective.
 
One of the economic policies implemented during the first two months following the first COVID-19 case in Chile, was the Employment Protection Law (EPL). The EPL allowed workers to use the funds accumulated in their individual unemployment insurance account as a new source of income while keeping their job contracts. This paper aims to measure the impact of the EPL on employment during the COVID-19 pandemic. 

It is important to know the effects of the EPL for several reasons. On the one hand, learning the effectiveness of the policy provide policymakers with valuable information to confront new negative shocks in the future.  If the EPL proves to be socially desirable, in the sense that its benefits outweighed the costs, then it could be part of the available tools that policymakers may use in time of big crises, not only COVID-19  driven shocks but any kind of negative shock.  On the other hand, since the EPL allows to use resources from individual accounts it is important to know whether it accomplishes its goal regarding employment maintenance. If it does not have a relevant effect, the policy would just translate into diminishing or emptying the accounts without any counterpart in employment. This would suggest that the policy could be seen as subsidies from the government to workers.    

Since the Employment Protection Law has national coverage when companies satisfy its requirements, there is not a natural group of firms for which the benefits of the EPL has been denied. Therefore, the assessment of the causal effect of this policy is not direct as there is not a clear control group. Elicitation of firms’ expectations allows to study the effect of the policy in an alternative way by comparing the expectations and realizations of those companies engaged in the EPL versus those that did not use it. This comparison allows to define an alternative way to measure the effect of the policy. In this sense, this paper is also related to the literature about measuring the accuracy of expectations using historical data. \cite{manski2004measuring} discusses different approaches to assess the accuracy of elicited expectations. One of them, and more closely related to the one used in this study, is the comparison between individual expectations and their realizations following respondents over time. Several sources have argued for this approach, see, for instance, \cite{katona1957federal, juster1966consumer, dominitz1998earnings, hurd2002predictive}. 

This paper is also connected to the literature about firms´ expectations and realizations. \cite{nerlove1983expectations} proposed some models of expectation formation and planning, and also analyzed their connection with realizations. \citep{nerlove1983expectations} used two business surveys where expectations, realizations and/or appraisals were analyzed over a set of aspects such as business conditions, demand, production, among others. In this context, realizations are not related to the actual actions taken by each individual firm, but rather to the actual outcome of variables at an aggregated level, such as demand and price levels. These factors  do not depend exclusively on decisions made by a firm individually. Therefore, deviations between expectations about the outcome of a variable at time $t$ and the actual realizations at time $t$ could be the product of factors that do not fully depend on the firm’s decision-making process. These factors include misinformation, poor quality of the expectation formation process, unseen impacts on industries, under/overreaction of other firms on shocks, among others. These deviations may affect the way in which the expectations will be formed if the firm is aware about this issue, but they will not necessarily impact the way in which decisions are made inside the firm.

Other authors also study expectations of firms. For example, \cite{coibion2015firms} studies the way firms build their expectations considering recent macroeconomic developments and how being uninformed affects firms' expectations of the future. \cite{buchheim2020firm} explores  firms' expectations as a determinant of managerial strategies during the COVID-19-19 crisis; in \cite{miyakawa2021firm} firms’ expectations for future sales are relevant as they affect the decision of exiting the market; \cite{koetse2006impact} investigate the impact of firms’ 2-year-ahead expectations and uncertainty on some key economic variables regarding their own levels of investment.

Our study is different to those previously cited since they do not analyze  firms' expectations to understand the factors that may affect the association between realization and expectations. When expectations and realizations are both related to the decision-making process of a firm, deviations between expectations and realizations are evidence of a change in the way the firm is planning to act. This change can be driven by internal processes or external shocks. In the case of the latter, a new public policy on about employment could affect the firm’s decisions about their own plan of firing a portion of their workers as a product of facing a negative external shock. This paper studies the way the EPL affects the relationship between firms´ expected and effective layoffs and, thus, it measures how this policy impacted firms´ decision making in order to reduce drops in employment.

The next section introduces the Employment Protection Law and presents some details that are relevant to understand how it works. Section \ref{sec:data} presents the data and descriptive statistics. Section \ref{sec:econometric} develops the empirical approach. Finally, section \ref{sec:Conclusions} concludes. 

\section{Employment Protection Law} \label{sec:EPL}

As a policy to protect job stability and decrease the negative impact of COVID-19-19 on income of formal workers, the law n°21,227 was enacted on the 1st of April 2020 to ``allow the use of unemployment insurance benefits in exceptional circumstances'' (\cite{EPL}), which is also known as the ``Employment Protection Law'' (EPL). The design of the EPL is considered special because, in normal conditions, the Chilean law does not permit temporal suspensions of job contracts in exceptional circumstances or emergencies. To improve some particular aspects of the original law, it was modified by the law n°21,232 (\cite{EPL2}), enacted on the 28th of May 2020. However, given the short distance in time between these two laws and the complementary character of the second, they are considered as one single employment policy, producing a common effect.

The EPL allows formal workers who meet some specific requirements to use the funds accumulated in their individual unemployment insurance account as a new source of income, keeping their contracts with their employers while the latter pay the corresponding costs associated with pension quotes and social security contributions of the workers while the period of suspension is on. The law works under certain situations: (a) temporary employment contract suspension by act of government authority (i.e. lockdown declaration), (b) temporary employment contract suspension agreement, and (c) temporary working  hours reduction agreement.

The suspension prevents companies from firing workers due to restrictions that temporarily affect their ability to generate income, and it allows employment relationships to be preserved for companies and workers, which can be resumed once the contingency is over\footnote{$https://www.bcentral.cl/documents/33528/2369613/rec\_impacto\_COVID-1919.pdf/373f37fc-7807-1e20-66e3-8d0c21b89db9?t=1593716468207$}.

\bigskip

\section{Data and Descriptive Statistics} \label{sec:data}

The information is obtained from monthly surveys addressed to a representative set of companies (public and private) and institutions from almost all economic sectors of the Chilean territory. 

Surveys are sent to more than 1,000 companies and institutions. These surveys aim to calculate two economic indicators called Remuneration Index (IR) and Labor Cost Index (ICMO). The data structure satisfies the longitudinal scheme needed to analyze expectations fulfillment since at least two observations are needed on the same subject in order to carry out this type of study. One of the observations is to elicit expectations and the other one is required to get their corresponding realizations.  
Companies self-report their expectations at month $t$ on whether they expect to fire workers within the following three months or not. Then, at month $t+3$ their actions reveal whether their expectation was met or not.\\
The question they answered is:\\

\bigskip

``Given the pandemic context of COVID-19-19, do you believe that your company will have to fire workers in the next three months?''\\

\bigskip
For which a yes/no answer is obtained as a response. Therefore, their expectation fulfillment assessment is binary, producing four possible outcomes depending on what they declared as expectation at $t$ and also on what their revealed action was at $t+3$.  Note that firms are the ones responsible for making the decision of firing some of their workers and whether they want to avoid it with the support of the EPL, meaning that the latter affects firms' decisions rather than individuals'. However, individuals' perception of job insecurity could also be reduced by the policy because of the positive association between individuals' expectation of job loss and unemployment rate, which is higher for people working in low wage occupations (see \cite{green2000job}).

Firms are asked about their expectations of firing some of their workers within three months following the interview and the response is a yes/no answer. These expectations can be not met in two ways. On the one hand, firms may expect to fire but do not do it later on. On the other hand, they may expect not to fire but do it anyway. The EPL aims to protect the employment source of workers, allowing them to access the benefits and supplements of the Unemployment Insurance, when some negative shocks -mainly associated to COVID-19- arise. 

We consider companies that did not use the EPL before declaring their expectations. The benefits of the EPL were not experienced by these companies at the moment of declaring whether they believe that, given the context of COVID-19-19, they will fire some of their employees in the next three months or not. Therefore, the use of the EPL after declaring their expectation is considered as one of the factors that could decrease the probability of expectation fulfillment for those who declared they will fire some employees in the next three months given the context of COVID-19-19. On the other hand, companies for which the EPL was already part of the set of information at hand at the moment of declaring their expectations were excluded from the analysis. Given that these companies already knew how the EPL worked and its benefits when building their expectations, the fact that the latter are fulfilled or not \textit{should be} driven by other factors different from the EPL itself.

Within the group of firms that did not use the EPL, 2.7\% of them expected to fire some of their workers in three months’ following the interview but finally did not do it, whereas this figure rises to 9.2\% within the group of firms that used the EPL. On the other side of the mismatch, 63.3\% of the firms that did not use the EPL had to fire some of their workers despite they did not expect to do so, while the corresponding figure within those who used the EPL was 36.3\%.

\pagebreak
\section{Econometric Specification} \label{sec:econometric}
Decisions made by companies regarding workers layoffs are observed as a discrete choice. According to \cite{mcfadden1973conditional}, these data may be used to make predictions on the choice made by the subject of analysis, in this case the firm. The model is set up to estimate the effects of the EPL on expectation fulfillment. We build a dummy variable $Y_{i,t}=1$ if the realizations of the $i$-th firm at months $t-2$, $t-1$ or $t$ are consistent with the elicited expectation at time $t-3$ and $Y_{i,t}=0$ otherwise. The econometric model $\forall i=1\ldots,n$ and $t=1,\ldots,T$ is
\begin{align}
    \text{logit}[P(Y_{i,t}=1|E_{i,t-3},EPL_{i,t},\mathbf{x}_{i,t})]=&\beta_0+ \beta_1 E_{i,t-3} + \beta_2 EPL_{i,t} \nonumber \\
    &+ \beta_3 E_{i,t-3}*EPL_{i,t} + \mathbf{x}'_{i,t}\boldsymbol{\delta}, \label{eq:model}
\end{align}
where
\begin{itemize}
    \item $E_{i,t-3}=1$ if, in $t-3$, the $i$-th firm expected to fire some of its workers in $t-2$, $t-1$ or $t$, and $0$ otherwise,
    \item $EPL_{i,t}=1$ if the $i$-th firm used the benefits of the EPL in $t$ and $0$ otherwise, and 
    \item the components of vector $\mathbf{x}_{i,t}$ describing the $i$-th firm at time $t$ are:
    \begin{itemize}
    \item $SIZE_{i,t}$ size of the $i$-th firm: small, medium, or large.
    \item $JS_{i,t-4}=1$ (job separation) if part of the $i$-th firm's workers stopped working for the firm, voluntarily or involuntarily, in $t-4$ and $0$ otherwise,
    \item $NW_{i,t}$ number of worker of the $i$-th firm at time $t$,
    \item $IND_{i,t}$ industry of the $i$-th firm, and
    \end{itemize}
\end{itemize}

We are mainly interested in the effect of $EPL_{i,t}$ on $P(Y_{i,t}=1)$. We included an interaction term between $E_{i,t-3}$ and $EPL_{i,t}$ to capture moderation effects depending on firms initial expectation regarding potential layoffs 3 months head. That is, we allow for the impact of the EPL to be different for those firms that reported they expected layoffs and those that did not expected to experiment layoffs. 

The effect of the EPL is
\begin{align}\label{eq:OR1}
    OR_{EPL}=\exp(\beta_2 + \beta_3 E_{i,t-3})
\end{align}
with
\begin{align*}
    OR_{EPL}\equiv \frac{\frac{P(Y_{i,t}=1|E_{i,t-3},EPL_{i,t}=1,\mathbf{x}_{i,t})}{1-P(Y_{i,t}=1|E_{i,t-3},EPL_{i,t}=1,\mathbf{x}_{i,t})}}{\frac{P(Y_{i,t}=1|E_{i,t-3},EPL_{i,t}=0,\mathbf{x}_{i,t})}{1-P(Y_{i,t}=1|E_{i,t-3},EPL_{i,t}=0,\mathbf{x}_{i,t})}}
\end{align*}
being the odds ratio between the odds of the expectation fulfilment given $EPL=1$ and the odds of the expectation fulfilment given $EPL=0$.\\

The effect of EPL on the probability of expectation fulfillment depends on whether companies declared they were expecting to fire some of their workers during the next three months or not. This distinction is captured by the interaction defined in model \eqref{eq:model} and equation \ref{eq:OR1}. As seen by equation \ref{eq:OR1} The parameters of main interest are $\beta_2$ and $\beta_3$ since they determine the effects of the policy for those firms who expected and did not expected to fire workers. The effect of the policy for those that did not expect to fire workers is:

\begin{align}\label{eq:OR_E0}
    OR_{EPL,E_{t-3}=0}=\exp(\beta_2)
\end{align}
The effect of the policy for those that expected to fire workers is:
\begin{align}\label{eq:OR_E1}
    OR_{EPL,E_{t-3}=1}=\exp(\beta_2+\beta_3)
\end{align}

On the one hand, we expect $\beta_2$ to be positive or zero since for those firms that did not expect to fire workers the effect of the policy could increase the expectation fulfillment. On the other hand, higher probabilities of expectation fulfillment mean firms are more likely to fire workers when they declared they expected to do it. Reducing this last effect is aimed by the EPL, that is, the marginal effect of the EPL to be negative for this group. Thus we would expect $\beta_2+\beta_3$ to be negative. 
\subsection{Results}

 Parameter estimates of model \eqref{eq:model} together with their corresponding standard errors and p-values are shown in Table \ref{table:logitResults}. 

When companies declared they were not expecting to fire some of their workers during the next three months ($E_{t-3}=0$), the odds ratio associated with the coefficient of $EPL_t$ is slightly greater than 1, 1.0042 ($exp(\widehat{\beta}_2)=exp(0.00419)$), which suggests there is no significant effect of EPL for these firms. This is confirmed by the fact that the coefficient associated with $EPL_t$ is not significant, indicating that the EPL does not produce a significant impact on the probability of expectation fulfillment for this type of companies. In fact, the average marginal effect of $EPL_t$ when $E_{t-3}=0$ on the probability of expectation fulfillment is slightly greater than zero, 0.05\% (see Table \ref{table:marginalEffects}). 

On the contrary, the p-value of the estimate associated with the interaction between $EPL_t$ and $E_{t-3}$ is low, 1.2\%. This indicates that there is a significant effect of $EPL_t$ on the probability of expectation fulfillment for companies that declared they were expecting to fire some of their workers during the next three months ($E_{t-3}=1$). The coefficient is negative and, therefore, the odds ratio is low too, 0.5 ($exp(\widehat{\beta}_2+\widehat{\beta}_3)=exp(0.00419-0.6970)$)), meaning that the EPL decreases the probability of expectation fulfillment when companies originally believed that they will fire some of their workers but finally did not do it. In fact, the average marginal effect of $EPL_t$ when $E_{t-3}=1$ on the probability of expectation fulfillment is almost -9.5\% (see Table \ref{table:marginalEffects}). 
Hence, the effect of $EPL_t$ is significant when companies did expect to fire workers and not significant when they did not. In order to analyze the overall significance of those parameter estimates associated with $EPL_t$ we tested $H_0: \beta_2=\beta_3=0$, and the resulting p-value was 0.005929. Therefore, the overall effect of $EPL_t$ is significant too.

\begin{table}
\begin{center}
\begin{tabular}{ l | r r r l }
 & Estimate & Std. Error & P-value &   \\ \hline 
Constant & -1.06800 & 0.201 & 1.06e-07 & *** \\ 
$E_{t-3}$ & 4.20300 & 0.104 & $<$ 2e-16 & *** \\ 
$EPL_{t}$ & 0.00419 & 0.175 & 0.981 &  \\ 
$E_{t-3}*EPL_{t}$ & -0.69700 & 0.278 & 0.012 & * \\ 
$Size:Medium_t$ & 0.32570 & 0.112 & 0.004 & ** \\ 
$Size:Small_t$ & 1.51400 & 0.100 & $<$ 2e-16 & *** \\ 
$JS_{t-4}$ & -1.42700 & 0.085 & $<$ 2e-16 & *** \\ 
$NW_{t}$ & -0.00002 & 2e-05 & 0.373 &  \\ 
$IND_{t}$ & - & - & - &   
\end{tabular}
\end{center}
\caption{Results of model \eqref{eq:model}. Estimates of 15 categories associated to different industries ($IND_{t}$) were omitted, from which only three were statistically significant at a level of 0.05 (Accommodation and food service, Manufacturing, and Mining).  \\Signif. codes:  0 ‘***’ 0.001 ‘**’ 0.01 ‘*’ 0.05 ‘.’ 0.1 ‘ ’ 1}
\label{table:logitResults}
\end{table}

Firm size, job separation and three industries are other variables whose parameter estimates are also significant. For firm size, the base category is large. Its parameter estimates show monotonic effects, indicating that the smaller the firm, the larger the probability of expectation fulfillment. In terms of average marginal effects, Table \ref{table:marginalEffects} shows that small firms increase their probability of expectation fulfilment in 11.9\% compared to large firms if they declared that they were expecting to fire workers. 
The corresponding figure for medium firms is 3.8\%. A factor that decreases these probabilities is the fact that part of the employees stopped working for firms the month before the one of the interview, voluntarily or involuntarily. This could be product of several reasons, for instance, at the moment of the interview firms might be experiencing some level of instability at least in terms of their human resources management and therefore their expectations were built with a high level of uncertainty, making the probability of meeting their expectations lower. Finally, there are three categories of the variable Industry that are significant. Its base category was chosen to be ``Water supply and waste management''. Given that it is related to basic services, it is one of the industries that should be less affected by COVID-19-19 in terms of their employment, and therefore a good point of reference for other industries. The three industries with a significant effect are ``Accommodation and food service'', ``Manufacture'', and ``Mining'' (see tables \ref{table:logitResults} and \ref{table:marginalEffects}). All of them with negative effects and from which ``Accommodation and food service'' is the one with the highest effect among industries. It is well known that this industry was one of the first negatively affected by the shock of the pandemic, rapidly reacting with a decrease in the number of workers, mainly because of the implications of breakdowns and also producing a high level of uncertainty, which could also be the reason why the probability of meeting their expectations is smaller than in the industry of reference.

\begin{table}
\begin{center}
\begin{tabular}{ l | r | r | r }
 & All & $E_{t-3}=0$ & $E_{t-3}=1$ \\ \hline
$E_{t-3}$ & 0.6845 & 0.6794 & 0.7052 \\
$EPL_{t}$ & -0.0184 & 0.0005 & -0.0946 \\
$Size:Medium_t$ & 0.0351 & 0.0344 & 0.0379 \\
$Size:Small_t$ & 0.2014 & 0.2218 & 0.1188 \\
$JS_{t-4}$ & -0.1789 & -0.1924 & -0.1243 \\
$NW_{t}$ & -0.0024 & -0.0024 & -0.0025 \\
$IND_{t}$ Accommodation and food service & -0.1063 & -0.1000 & -0.1315 \\
$IND_{t}$ Manufacture & -0.0652 & -0.0642 & -0.0690 \\
$IND_{t}$ Mining & -0.0655 & -0.0645 & -0.0694
\end{tabular}
\end{center}
\caption{Average marginal effects for different subsets of data. Other industries omitted as their corresponding parameter estimates are not significant at a level of 0.05 (reference category: Water supply and waste management).}
\label{table:marginalEffects}
\end{table}

Table \ref{table:profiles} shows the estimated probabilities of expectation fulfillment for a group of companies depending on their size and on whether they used the EPL or not during the next three months after they declared that they expected to fire some of their workers in the same period. In general, those estimated probabilities are smaller when comparing companies that used the EPL against those that did not use it, meaning that they did not fire some of their workers and therefore their expectations were not met. This effect is greater for large companies, reaching a decrease of 14.7 percentage points, whereas for small companies the corresponding decrease is of 5.7. Although the effect of EPL on the estimated probabilities is the smallest for small companies compared to others, its effect is still highly significant, as shown in Table \ref{table:logitResults} (p-value $<$ 2e-16). These results suggest that for small companies it is more difficult to overcome negative shocks even with the support of an employment policy such as the EPL, probably because they are more exposed to other factors regarding their decision of firing some of their workers. On the more positive side of the policy, it turned out to be a more effective policy for large and medium-sized companies, because it reduced more than 12 percentage points the probability of fulfilling their expectation of firing some of their workers.

\begin{table}
\begin{center}
\begin{tabular}{ l | c c | c }
 \multirow{2}{*}{Company size} & \multicolumn{2}{c|}{EPL} & \multirow{2}{*}{Difference (1-0)} \\
 & 0 & 1 &    \\ \hline 
Large & 0.761 & 0.614 & -0.147 \\ 
Medium & 0.815 & 0.688 & -0.127 \\ 
Small & 0.935 & 0.879 & -0.057 
\end{tabular}
\end{center}
\caption{Estimated probabilities of expectation fulfillment at time $t$ for companies that, at time $t-3$, believed that they will fire some of their workers at time $t-2$, $t-1$ or $t$. Companies of the manufacturing industry, with an average number of 906.4 workers, and experienced job separations at time $t-4$.}
\label{table:profiles}
\end{table}

Results suggest that for those firms that expected to fire
workers, the probability they finally ended up with a job separation
is 50% of those that did not used the EPL. 

\subsection{Discussion} \label{subsec:discussion}

Equation \eqref{eq:model} shows that the effect of the EPL is coming from the comparison of the probability of $Y=1$ between both groups, that is, for the group for which $EPL=1$ and the group for which $EPL=0$:
\begin{align}
     Pr(Y_{i,t}=1|E_{i,t-3},EPL_{i,t}=1) - Pr(Y_{i,t}=1|E_{i,t-3},EPL_{i,t}=0) \label{eq:effectProb}
\end{align}

When conditioned on the group for which $E_{i,t-3}=1$, the probability $Pr(Y_{i,t}=1)$ is the probability of a job separation. Section \ref{sec:data} showed that job separations could be due to layoffs, voluntary quits, among others and it is impossible to distinguish among all types of separation. Thus, the layoff probability is overestimated. However, as stated in \eqref{eq:effectProb}, identification of the effect of the policy is coming from the comparison of the probability for both groups, so if we could assume that the EPL policy did not affect the distribution of the job separation rate (among layoffs, quits, others) the estimation of the effect would not be biased. To the extent that we could assume companies that used the EPL policy are similar in terms of the proportions of types of job separation, then we can estimate the effect of the EPL policy through equation \eqref{eq:model}.This assumption is violated, for instance, if workers in firms that adopted the EPL tend to present less voluntary quits when compared with workers in firms that did not adopt the EPL policy. In that case, estimates in equation \eqref{eq:model} would underestimate the effect of the policy and, thus, they could be interpreted as a lower bound of the effect of the policy.
\section{Conclusions} \label{sec:Conclusions}

The COVID-19 pandemic brought about a negative shock to the whole economic activity, specially on employment. Many non conventional economic policies came up to overcome these ill effects. We study the effects of the Employment Protection Law, one of those non traditional policies, on expectation fulfilment as an alternative way to assess the impact of a novel economic policy in times of the COVID-19 pandemic. 

We took advantage of the firms self reported expectation about layoffs 3 months ahead in order to analyze whether their expectation finally fulfilled. 

We run a logit model to analyze the impact of the EPL on the probability of expectation fulfilment. We included and interaction term to consider moderation effects that could be present because of firms initial expectations regarding potential layoffs 3 months head: those who reported that expected layoffs and those that did not expected to experiment layoffs during the following 3 months. 
We found an effect for those companies that were expecting to fire some of its workers during the following three months. For them, the probability of having experimented a job separation was reduced by half in comparison with companies that did not enrolled in the EPL.

\bibliographystyle{apalike}

\end{document}